\documentclass[Physsubmission, Phys]{SciPost}

\hypersetup{
    colorlinks,
    linkcolor={red!50!black},
    citecolor={blue!50!black},
    urlcolor={blue!80!black}
}

\usepackage[bitstream-charter]{mathdesign}
\DeclareSymbolFont{usualmathcal}{OMS}{cmsy}{m}{n}
\DeclareSymbolFontAlphabet{\mathcal}{usualmathcal}

\begin{document}

\begin{center}{\Large \textbf{
Resummation for $tqH$ production\\
}}\end{center}

\begin{center}
Matthew Forslund\textsuperscript{1} and
Nikolaos Kidonakis\textsuperscript{2}
\end{center}

\begin{center}
{\bf 1} Stony Brook University, Stony Brook, NY 11794, USA
\\
{\bf 2} Kennesaw State University, Kennesaw, GA 30144, USA
\\
* matthew.forslund@stonybrook.edu
\end{center}

\begin{center}
\today
\end{center}

\definecolor{palegray}{gray}{0.95}
\begin{center}
\colorbox{palegray}{
  \begin{tabular}{rr}
  \begin{minipage}{0.1\textwidth}
    \includegraphics[width=22mm]{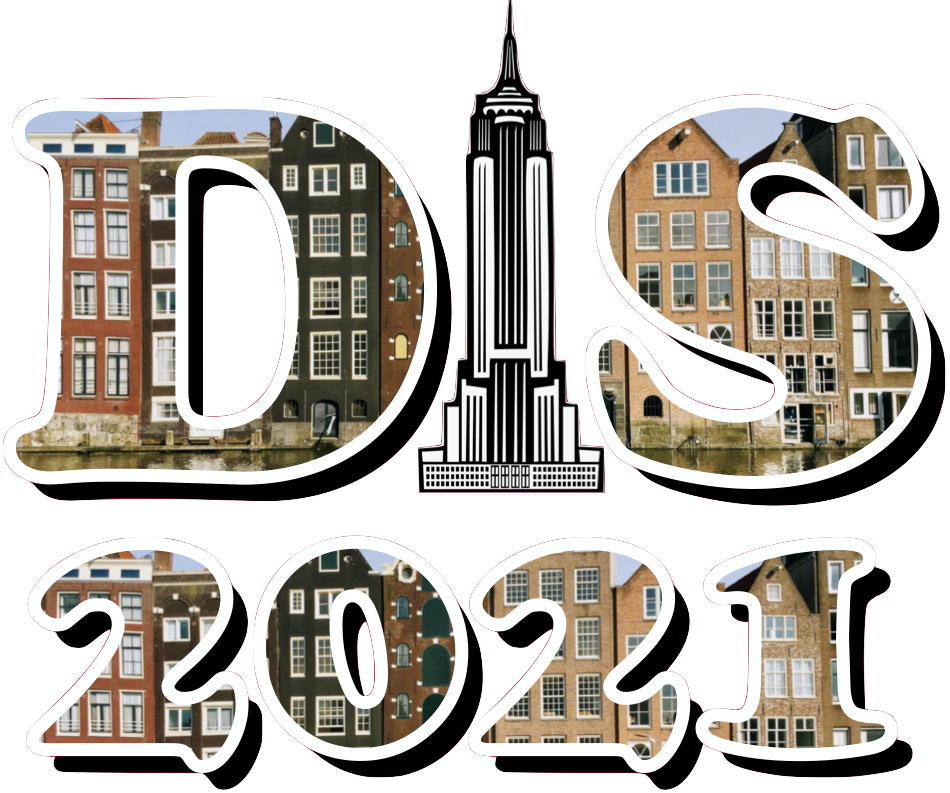}
  \end{minipage}
  &
  \begin{minipage}{0.75\textwidth}
    \begin{center}
    {\it Proceedings for the XXVIII International Workshop\\ on Deep-Inelastic Scattering and
Related Subjects,}\\
    {\it Stony Brook University, New York, USA, 12-16 April 2021} \\
    \doi{10.21468/SciPostPhysProc.?}\\
    \end{center}
  \end{minipage}
\end{tabular}
}
\end{center}

\section*{Abstract}
{\bf
We present results with soft-gluon resummation for the associated production of a single top quark and a Higgs boson. We present analytical results for the higher-order soft-gluon corrections and numerical results for the total cross section and top-quark transverse momentum and rapidity distributions at LHC energies.
}

\section{Introduction}
\label{sec:intro}

\begin{figure}[b]
\centering
\includegraphics[width=0.7\textwidth]{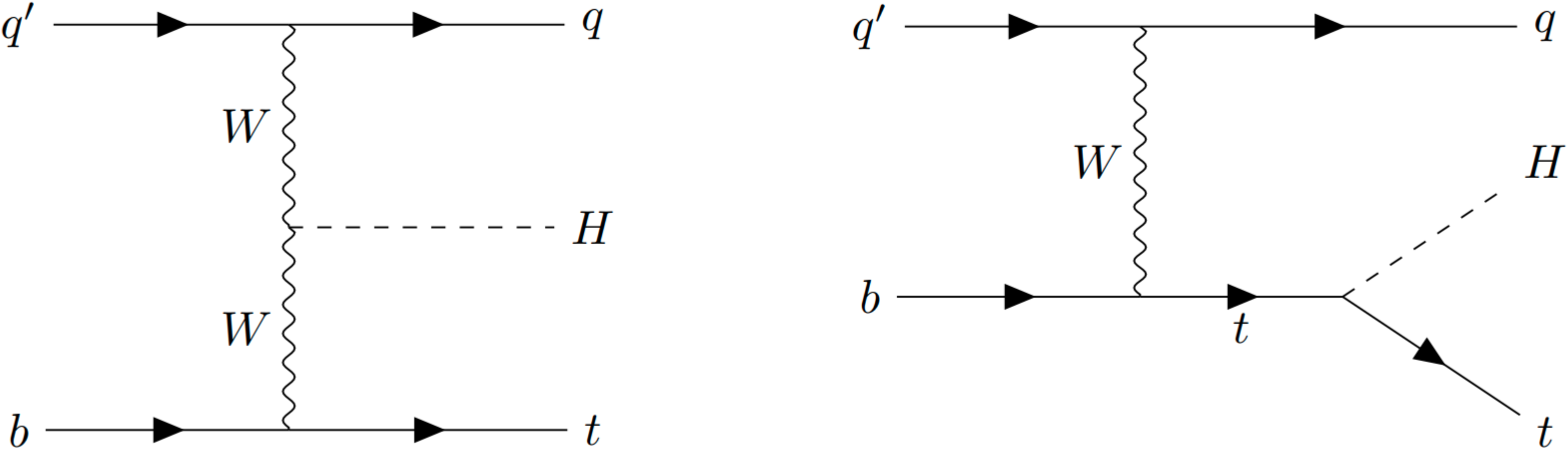}
\caption{Leading-order diagrams for $tqH$ production}
\label{diagrams}
\end{figure}

Processes at current and future colliders involving the Higgs boson and the top quark are of central importance for precision determination of standard model parameters and for searches for new physics. The associated production of a single top quark with a Higgs boson is of particular importance for the determination of the coupling of the Higgs to the $W$ boson and of the sign and magnitude of the top Yukawa coupling, with experimental searches having been made at the LHC at 8 TeV \cite{CMS8tev} and 13 TeV \cite{CMS13tev}. The tree-level Feynman diagrams for the process are shown in Figure \ref{diagrams}.

QCD calculations at next-to-leading-order (NLO) for $tqH$ production have been performed in \cite{CER,DMMZ} and the corrections were shown to be sizeable, on the order of 10\% at LHC energies. Since the corrections are large, it is important to go to higher orders for accurate theoretical predictions to complement experimental searches.

For many $2\rightarrow2$ processes involving top quarks, the full QCD corrections have been found to be numerically dominated by soft-gluon corrections at LHC and even higher energies (see \cite{NKtoprev} for a review). Soft-gluon corrections can therefore often be thought of as excellent approximations to full results. We find this is also the case for $tqH$ production at LHC energies \cite{FK2021}.

The form of the threshold variable appearing in the soft-gluon corrections depends on the kinematics. The soft-gluon resummation formalism in single-particle-inclusive (1PI) kinematics was recently extended to processes with an arbitrary number of final state particles \cite{FK2020}. $tqH$ production is the first application of this formalism to a $2\rightarrow3$ process and serves as a test case for the validity of soft-corrections as an approximation to full results for processes beyond $2\rightarrow2$.

\section{Soft-gluon resummation}

Here we briefly describe the corrections as obtained in \cite{FK2021,FK2020}. We have the partonic process $b(p_b)+q'(p_{q'}) \rightarrow t(p_1) + q(p_2) + H(p_3)$. We define the kinematic invariants $s = (p_b+p_{q'})^2$, $t=(p_b-p_1)^2$, $u=(p_{q'}-p_1)^2$, and $p^2_{23} = (p_2+p_3)^2$. With extra gluon emission, momentum conservation requires $p_b+p_{q'}=p_1+p_2+p_3+p_g$. We can then define a threshold variable $s_4 = s+t+u-m_t^2-p_{23}^2 = (p_2+p_3+p_g)^2-(p_2+p_3)^2$, in analogy with the $2\rightarrow2$ definition, which clearly goes to zero as $p_g\rightarrow0$ and carries the physical meaning of the extra energy from soft-gluon emission.

We can then take Laplace transforms of the partonic cross section using this threshold variable as ${\tilde \sigma}(N)=\int_0^s (ds_4/s) \,  e^{-N s_4/s} \, \sigma(s_4)$ where $N$ is the transform variable. The cross section then factorizes \cite{NKtoprev,FK2021,FK2020,NKGS1,NKGS2,KOS,LOS} as
\begin{equation}
    {\tilde \sigma}_{bq' \to tqH}(N)= {\psi}_b(N_b) \, {\psi}_{q'}(N_{q'}) \, {J_q} (N) \, \, {\rm tr}\left\{H_{bq' \to tqH} \, {S}_{bq' \to tqH}\left(\frac{\sqrt{s}}{N \mu_F} \right)\right\} \, ,
\label{sigN}
\end{equation}
where $\psi_{b,q'}$ and $J_q$ are distributions describing collinear emission from the initial state partons $b,q'$ and the final state light quark $q$, respectively \cite{S1987}. Both $H_{bq' \to tqH}$ and $S_{bq' \to tqH}$ are $2\times2$ matrices in color space specific to this process, where $H_{bq' \to tqH}$ is an $N$-independent hard-scattering function and $S_{bq' \to tqH}$ is a soft function describing noncollinear soft-gluon emission. 

The $N$ dependence of $S_{bq' \to tqH}$ is resummed via renormalization-group evolution, as it obeys the relation
\begin{equation}
\left(\mu_R \frac{\partial}{\partial \mu_R}
+\beta(g_s)\frac{\partial}{\partial g_s}\right) { S}_{bq' \to tqH}
=-(\Gamma_{\! S \, bq' \to tqH})^{\dagger} \; { S}_{bq' \to tqH}-{ S}_{bq' \to tqH} \; \Gamma_{\! S \, bq' \to tqH}
\end{equation}
where $\Gamma_{\! S \, bq' \to tqH}$ is the soft anomalous dimension matrix which governs the evolution of the soft function and yields the exponentiation of logarithms of $N$ in the resummed cross section.

The resummed cross section is then expanded to fixed order and inverted back into momentum space to obtain the soft-gluon corrections involving logarithmic plus distributions of the threshold variable $\mathcal{D}_k(s_4) = [(\ln^k(s_4/m_t^2))/s_4]_+$. The NLO soft-gluon corrections for $bq'\to tqH$ are then given by
\begin{align}
E_1\frac{d{\hat{\sigma}}_{bq' \to tqH}^{(1)}}{d^3p_1}=& F^{LO}_{bq' \to tqH} \frac{\alpha_s(\mu_R)}{\pi} \left\{c_3 \, {\cal D}_1(s_4) + c_2 \, {\cal D}_0(s_4) +c_1 \, \delta(s_4)\right\} \nonumber \\ &+\frac{\alpha_s(\mu_R)}{\pi} \, A_{bq' \to tqH} \, {\cal D}_0(s_4) \, ,
\label{aNLO}
\end{align}
where $F^{LO}_{bq' \to tqH}= {\rm tr}\{H_{bq' \to tqH}^{(0)} \, S_{bq' \to tqH}^{(0)}\}$ involves the leading-order (LO) cross section and \linebreak $A_{bq' \to tqH}={\rm tr} \left\{H_{bq' \to tqH}^{(0)} \, \Gamma_{\! S \, bq' \to tqH}^{(1) \, \dagger} \, S_{bq' \to tqH}^{(0)}
+H_{bq' \to tqH}^{(0)} \, S_{bq' \to tqH}^{(0)} \, \Gamma_{\! S \, bq' \to tqH}^{(1)}\right\}$. The coefficients are given by $c_3=3C_F$,
\begin{equation}
    c_2=-2 \, C_F \, \ln\left(\frac{(p^2_{23}-u)(p^2_{23}-t)}{m_t^4}\right) 
-\frac{3}{4}C_F -3 \, C_F \ln\left(\frac{m_t^2}{s}\right)
-2 \, C_F \ln\left(\frac{\mu_F^2}{m_t^2}\right)
\label{c2}
\end{equation}
\begin{equation}
    c_1=C_F\left[\ln\left(\frac{(p^2_{23}-u)(p^2_{23}-t)}{m_t^4}\right)
-\frac{3}{2}\right]\ln\left(\frac{\mu_F^2}{m_t^2}\right)  \, .
\label{c1mu}
\end{equation}

The NNLO soft corrections are much more complicated and follow directly from expressions given in \cite{FK2020}. 

\section{Results}

We now present numerical results for total $tqH+\bar{t}qH$ production as discussed in more detail in \cite{FK2021}. Throughout we use $m_H=125$ GeV, $m_t = 172.5$ GeV and MSHT20 pdf sets \cite{MSHT20}, although we note that different pdf choices have little impact on these results. We set the renormalisation and factorization scales to be equal throughout ($\mu = \mu_F = \mu_R$). In these results we refer to the NLO soft corrections added to the leading-order result as approximate next-to-leading-order (aNLO), the exact next-to-leading-order results as NLO, and the sum of the exact NLO and NNLO soft corrections as aNNLO. The soft results are computed from resummation at next-to-leading-logarithm accuracy.

\begin{figure}[h]
\centering
\includegraphics[width=0.49\textwidth]{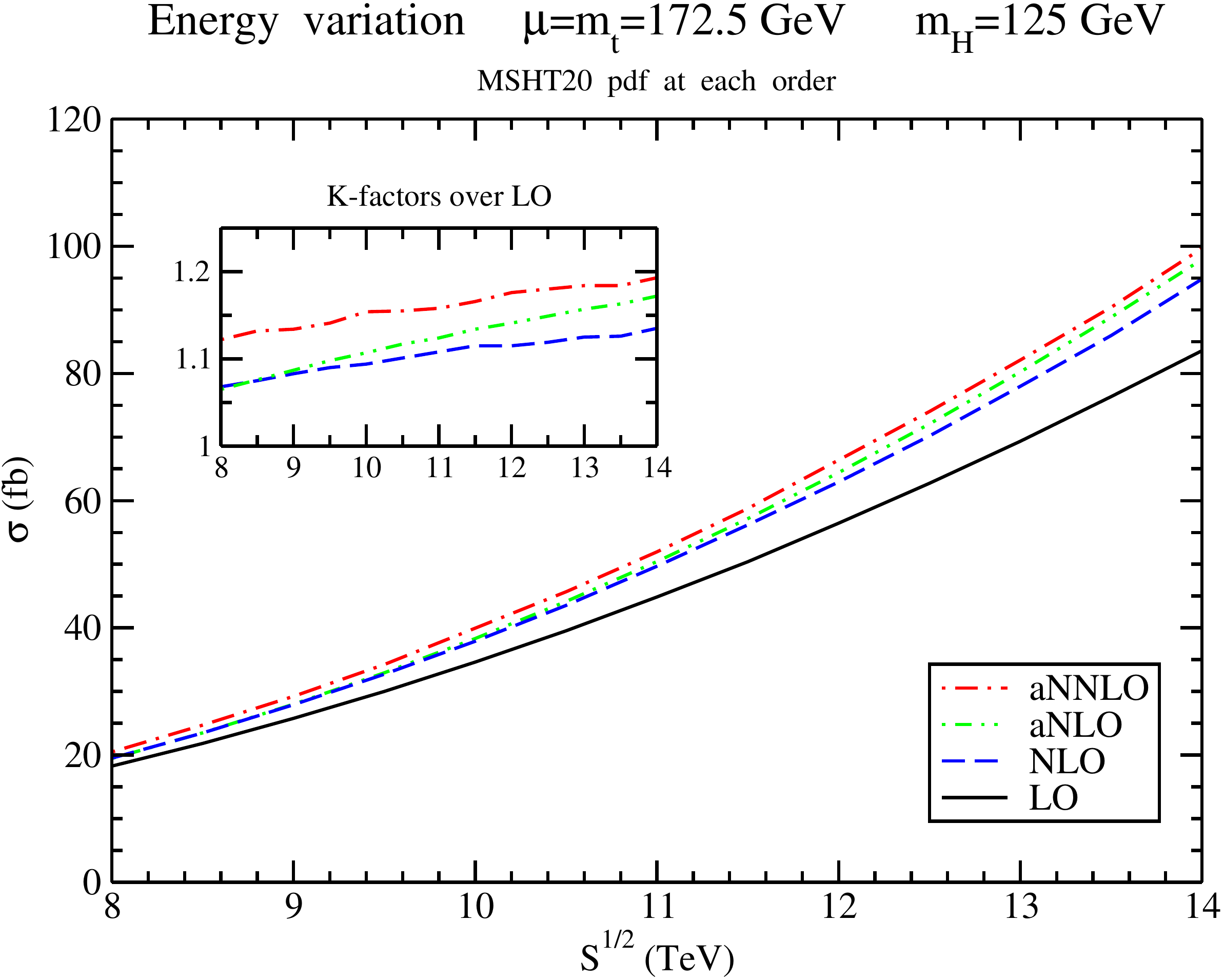}
\includegraphics[width=0.49\textwidth]{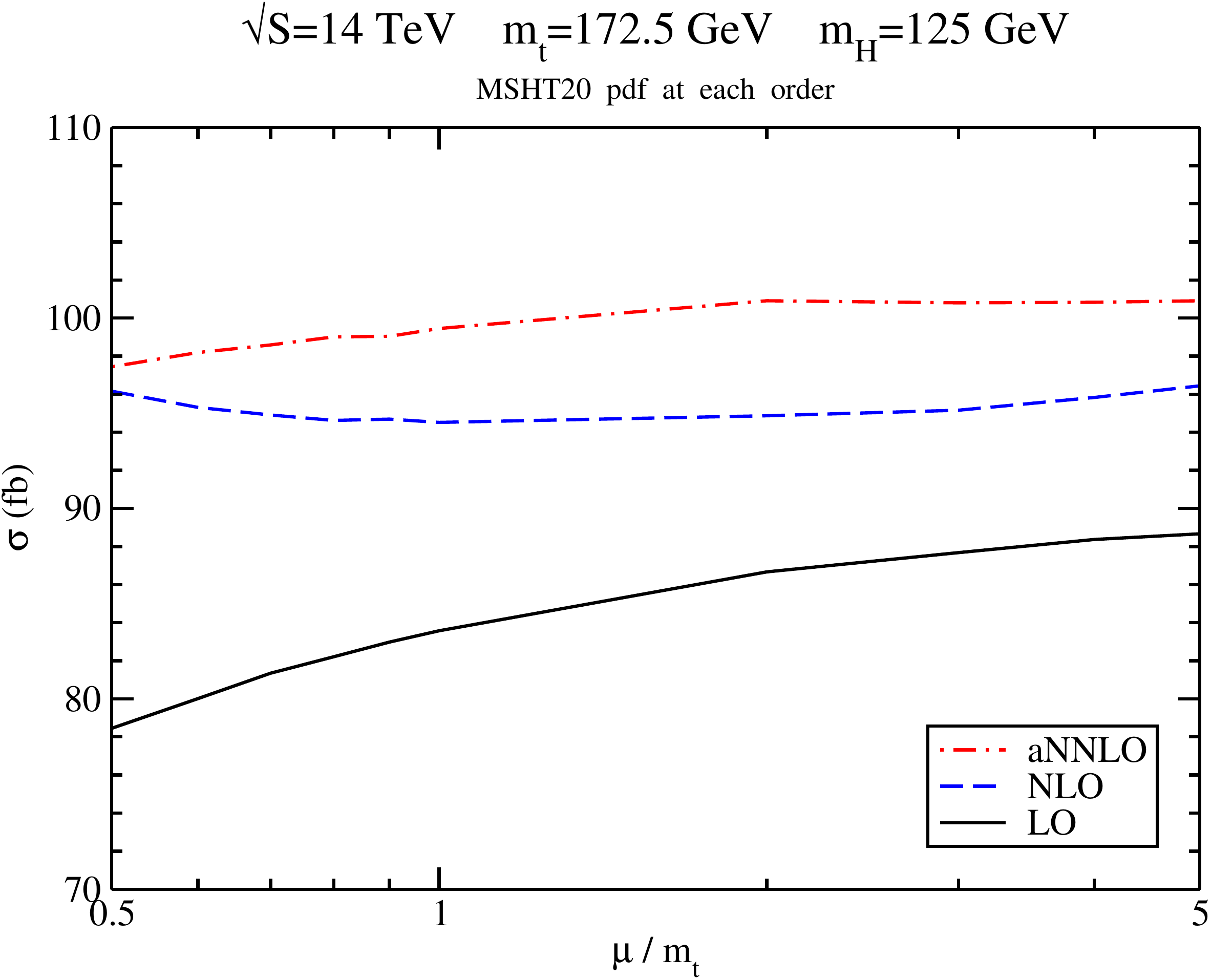}
\caption{The total $tqH + \bar{t}qH$ cross section. (left) Energy dependence of the cross section at $\mu=m_t$. (right) Scale dependence of the cross section at 14 TeV.}
\label{crosssection}
\end{figure}

We first look at the total cross section as a function of energy in the left plot of Figure \ref{crosssection}. An inset plot shows the K-factors over the LO result for the corrections. Here we can see a direct comparison between the aNLO results and the exact NLO, finding that they are quite close for the entire range. At 8 TeV the difference is completely negligible; at 14 TeV the difference is larger, although the NLO result is still completely within the scale variation range of the aNLO. The aNLO can therefore be thought of as a very good approximation to the exact result for LHC energies. At aNNLO, the corrections result in a substantial increase in the cross section over LO, nearly 20\% at 14 TeV.

The right plot of Fig. \ref{crosssection} shows the scale variation of the cross section at 14 TeV between $\mu=0.5m_t$ and $\mu=5m_t$ as an indication of the theoretical uncertainty. The result at aNNLO is very stable with a $\sim$3.5\% variation over the entire range, far smaller than at LO, as expected for higher order corrections. Similar results hold for the rest of the energy range in the left plot.

\begin{figure}[h]
\centering
\includegraphics[width=0.49\textwidth]{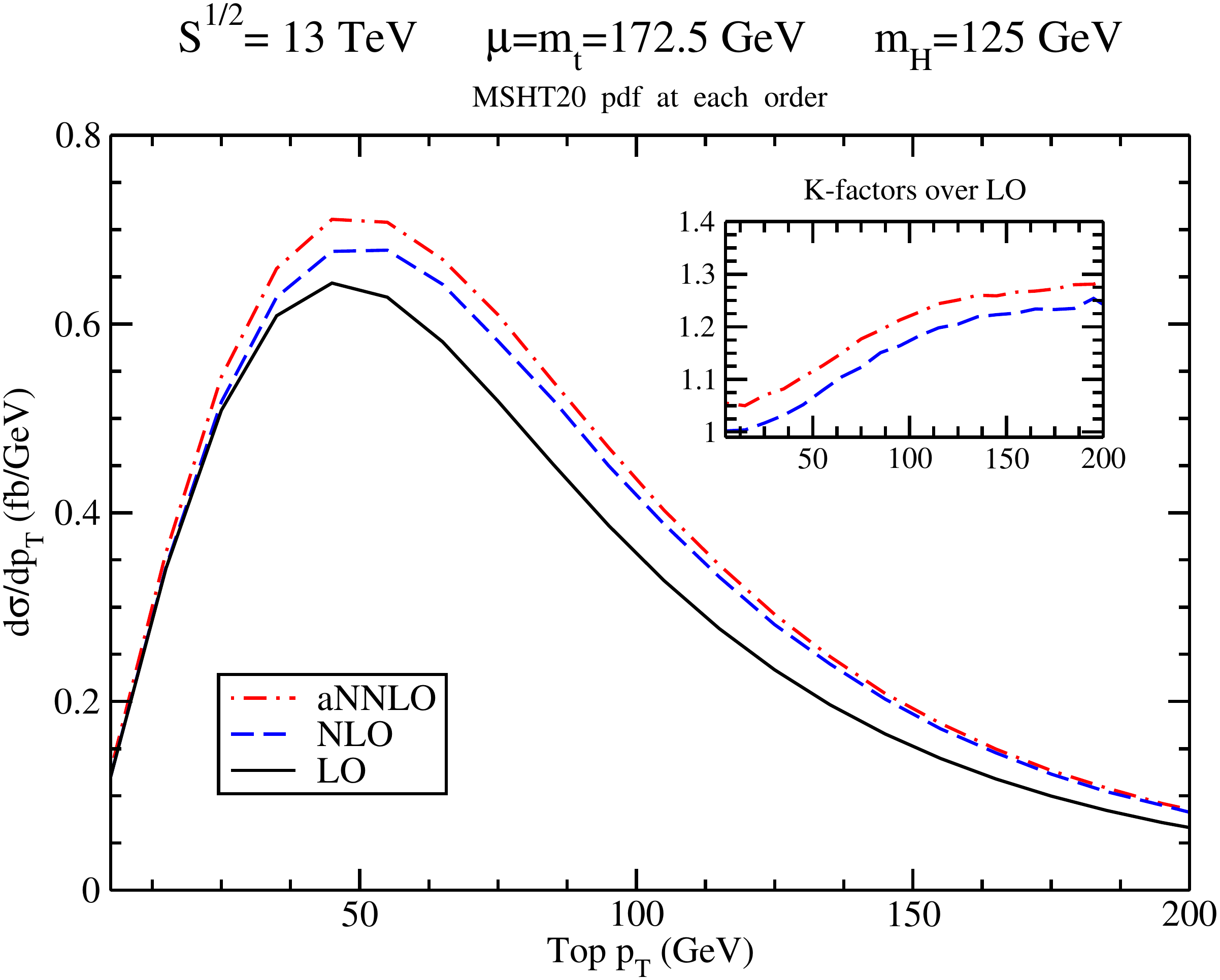}
\includegraphics[width=0.49\textwidth]{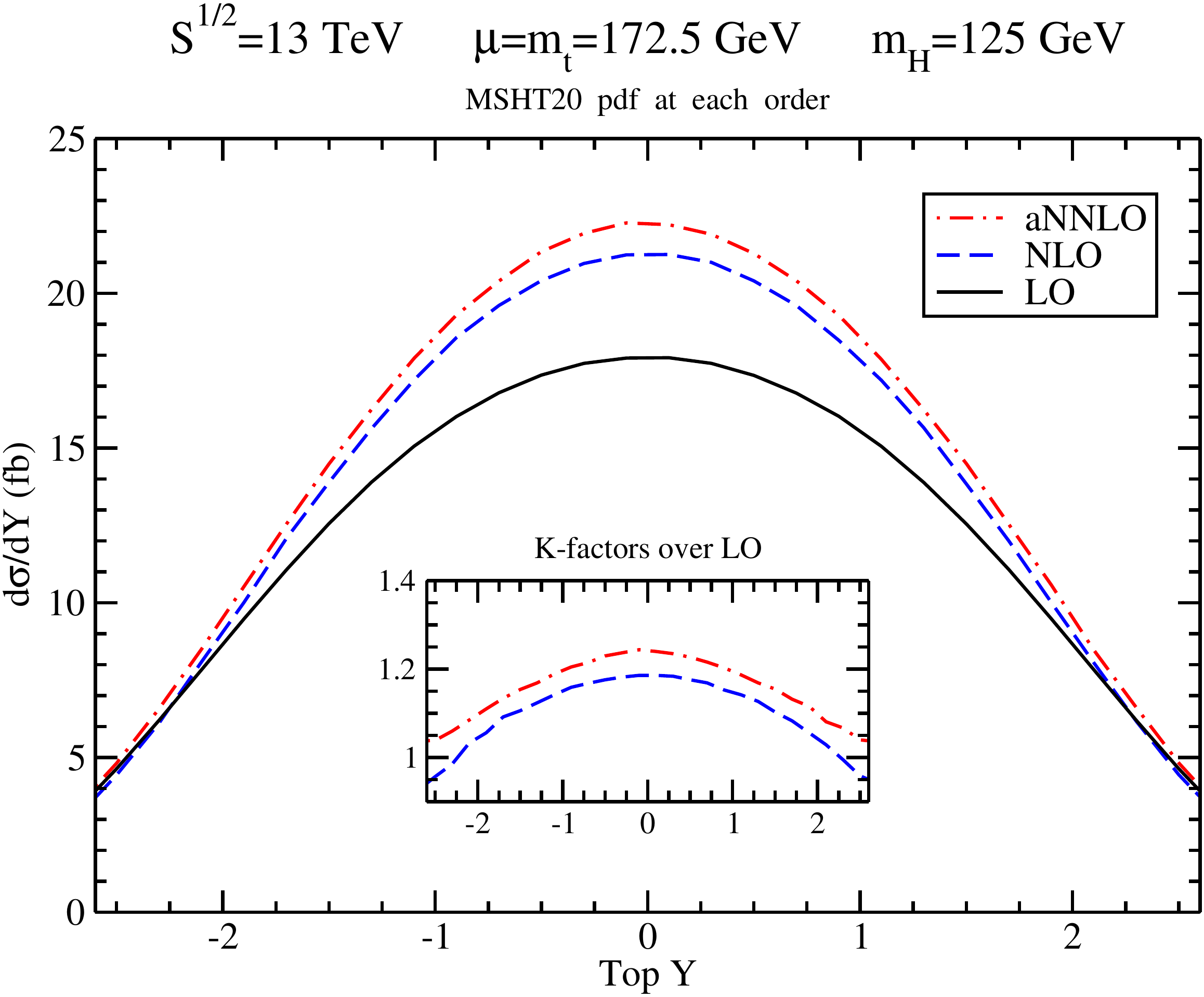}
\caption{Top-quark (left) transverse-momentum distributions and (right) rapidity distributions at LO, NLO, and aNNLO. Inset plots show K-factors relative to LO.}
\label{diffdists}
\end{figure}

Our formalism also allows us to compute 1PI differential distributions, which can be more sensitive to deviations due to new physics. Figure \ref{diffdists} shows top-quark differential distributions for total $tqH+\bar{t}qH$ production, in both transverse momentum and rapidity, at 13 TeV. The combined NLO and aNNLO corrections are substantial for both distributions, up to a $\sim$30\% enhancement at high $p_T$ and a $\sim$25\% enhancement at zero rapidity over LO. Distributions at 14 TeV, as well as a more detailed comparison of the aNLO and exact NLO results at the level of differential distributions can be found in \cite{FK2021}. The results at 14 TeV are very similar to those shown here, and the NLO soft corrections remain an excellent approximation to the full result even in differential distributions.

\section{Conclusion}
We have presented results for soft-gluon resummation for $tqH$ production, the first $2\rightarrow3$ application in 1PI kinematics. We have shown that our results approximate well the exact results for LHC energies at NLO. We have presented numerical results at aNNLO including total $tqH+\bar{t}qH$ cross sections and top-quark differential distributions. The aNNLO corrections are shown to be significant, and their inclusion improves the theoretical predictions.

\paragraph{Funding information}
This material is based upon work supported by the National Science Foundation under Grant No. PHY 1820795.

\nolinenumbers

\end{document}